\documentclass[aps,superscriptaddress,prb,floatfix,twocolumn,preprintnumbers,amsmath,amssymb]{revtex4}

\usepackage{graphicx}
\usepackage{dcolumn}
\usepackage{bm}

\begin{document}


\title{Phase Transitions in High Purity Zr Under Dynamic Compression}

\author{C. W. Greeff}
 \email{greeff@lanl.gov}
\affiliation{Los Alamos National Laboratory, Los Alamos, New Mexico 87545}

\author{J. Brown}
\affiliation{Sandia National Laboratories, Albuquerque, NM 87185}

\author{N. Velisavljevic}
\affiliation{High Pressure Collaborative Access Team (HPCAT), Advanced Photon 
Source, Argonne, IL 60439, and Physics Division, Lawrence Livermore National 
Laboratory, Livermore, CA 94550}

\author{P. A. Rigg}
\affiliation{Dynamic Compression Sector, Institute for Shock Physics,
Washington State University, Argonne, IL 60439}

\date{\today}

\begin{abstract}
We present results from ramp compression experiments on high-purity Zr
that show the $\alpha \rightarrow \omega$,  $\omega \rightarrow \beta$, 
as well as reverse $\beta \rightarrow \omega$ phase transitions.
Simulations with a multi-phase equation of state and phenomenological 
kinetic model match the experimental
wave profiles well. While the dynamic $\alpha \rightarrow \omega$ transition
occurs $\sim 9$~GPa above the equilibrium phase boundary, 
the $\omega \rightarrow \beta$  transition 
occurs within 0.9~GPa of equilibrium. We estimate that the dynamic compression
path intersects the equilibrium $\omega - \beta$ line at $P= 29.2$~GPa,
and $T = 490$~K. The thermodynamic 
path in the interior of the sample lies $\sim 100$~K above the isentrope
at the point of the $\omega \rightarrow \beta$ transition. Approximately 
half of this dissipative temperature rise is due to plastic work, and half
is due to the non-equilibrium $\alpha \rightarrow \omega$ transition.
The inferred rate of the $\alpha \rightarrow \omega$ transition is several
orders of magnitude higher than that measured in dynamic diamond anvil cell
(DDAC) experiments in an overlapping pressure range. We discuss a model for 
the influence of shear stress on the nucleation rate.
We find that the shear stress $s_{ji}$ has the same effect on the 
nucleation rate 
as a pressure increase 
$\delta P = c \epsilon_{ij} s_{ji}/(\Delta V/V),$
where $c$ is a geometric constant $\sim 1$ and, $\epsilon_{ij}$ are the
transformation shear strains. The small fractional volume change 
$\Delta V/V \approx 0.1$ at the $\alpha \rightarrow \omega$ transition
amplifies the effect of shear stress, and we estimate that for this case
$\delta P$ is in the range of several GPa. Correcting our transition rate
to a hydrostatic rate brings it approximately into line with the DDAC results,
suggesting that  shear stress plays a significant role in the 
transformation rate.

\end{abstract}

\maketitle
\section{Introduction}               

Metallic Zr and its alloys have practical applications in chemical processing
and nuclear power.\cite{northwood1985} The high pressure properties 
and phase diagram of pure Zr have been studied extensively 
\cite{greeff_ti_zr_sccm,greeff_zr_eos,liu_zr_ult_08,dewaele_a-w_xafs_16,
zilbershteyn_aw_ti_zr_73,zhang_zr_phase_diag,akahama91,xia90,xia91,
ono_omega_beta_15,cerreta2005,prigg_zr_jap,jacobsen15,errandonea2005,
zhao_thermal_eos_2005}
using both static and dynamic compression techniques. The high pressure
phase diagram is shown in figure \ref{phase_diag}. The ambient
pressure $\alpha$ phase has the hcp structure. The $\beta$ phase, with
bcc structure appears at high temperature and high pressure. The
$\omega$ phase, with a hexagonal structure with 3 atoms per cell, occupies
intermediate $T$ and $P$. The sequence of phases with increasing pressure
on the room $T$ isotherm, isentrope, and Hugoniot is $\alpha-\omega-\beta$.
Dynamic compression studies \cite{lasl_shock_comp,cerreta2005,prigg_zr_jap,gorman_laser_recovery_2020,radousky_melting_2020,kalita_dxrd_2020} 
have for the 
most part been focused on shock compression to measure the Hugoniot
and investigate the $\alpha-\omega$, $\alpha-\beta$, and melting transitions.

In shock compression, an  abrupt shock wave is driven through the sample,
typically by a high velocity impact. In contrast, ramp 
compression \cite{hawke_isentropic_comp_1972,hall_isentropic_2001} 
is achieved by
a smoothly varying pressure wave. Ramp compression is less dissipative than
shock compression, allowing investigation of the equation of state (EOS)
and phase transitions at lower temperatures than shock loading.
Under shock loading, a phase transition will not be evident in the wave
profile if the shock pressure is too high so that the transition is overdriven.
This is illustrated in figure 3 of ref. \onlinecite{prigg_zr_jap}.
Under ramp compression, transitions do not become overdriven. Any
phase change encountered on the compression path should leave an
imprint on the wave profile.

Here we present new ramp compression data on Zr, obtained via magnetic
drive on the Z machine. We employ simulations using a multi-phase
equation of state \cite{greeff_zr_eos} together with a phenomenological 
kinetic model \cite{greeff_dyn_phase_2016} to interpret these experiments
together with shock loading data.\cite{prigg_zr_jap}
Our simulations agree well with experimental velocity profiles $v(t)$ measured
at the sample window interface for both ramp and shock compression.
Consistent with earlier work\cite{prigg_zr_jap,greeff_ti_zr_sccm} we find 
strong kinetic effects on the $\alpha-\omega$ transition. The non-equilibrium 
transition contributes significantly to the dissipative heating during ramp 
compression. The Z-machine data shows the higher pressure $\omega-\beta$ 
transition in both the forward and reverse directions. 
This transition occurs closer to equilibrium than the $\alpha-\omega$ 
transition. The presence of the 
forward and reverse transitions allows us to refine the equilibrium 
phase diagram. Comparing the $\alpha-\omega$ transformation rates inferred 
from dynamic compression with those measured in a dynamic diamond anvil cell
(DDAC)\cite{jacobsen15} shows the dynamic compression rate to be orders of 
magnitude higher that that in the DDAC in an overlapping pressure range.
We consider a model for the influence of shear stress on the nucleation
rate, and find that it is of the correct magnitude to explain the
difference.

\section{Experiments}

\begin{figure*}
\includegraphics[width=01.0\textwidth]{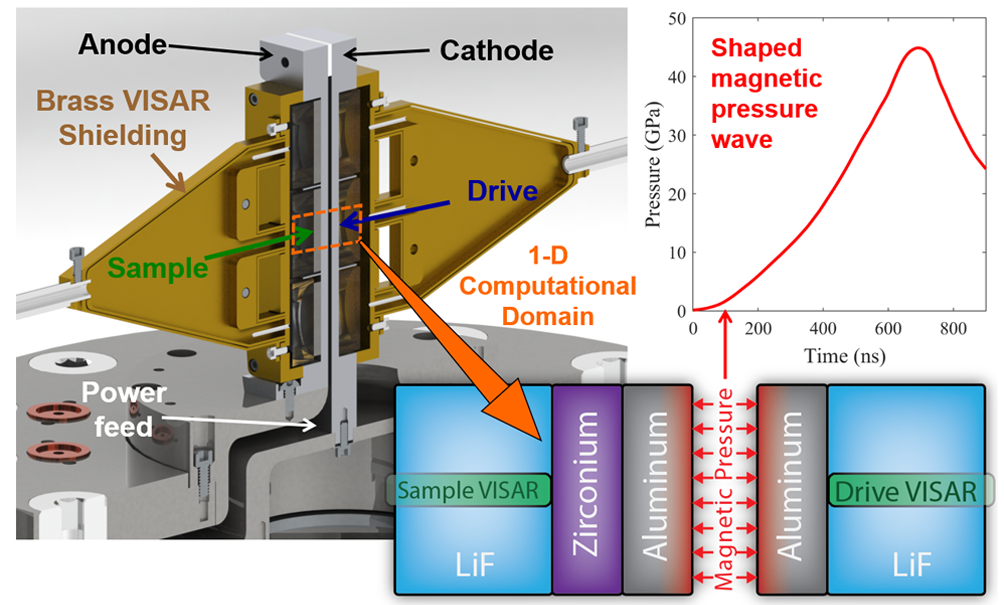}
\caption{\label{fig_RampConfig} Configuration for the magnetic ramp compression experiments. Each drive/sample pair is analyzed independently through a set 1-D simulations where the magnetic pressure boundary condition is inferred from the drive measurement which then enables simulations of the zirconium sample side.}
\end{figure*}

The experimental configuration for the ramp compression experiment, Z2913, is summarized in Figure \ref{fig_RampConfig}. The geometry shown is referred to as a stripline \cite{Lemke2011} and consists of two parallel aluminum electrodes which are 2 mm thick separated by a gap of 1 mm. From top to bottom, three Zr samples with thicknesses of 1.01, 1.25, and 1.51 mm are glued to the anode electrode with angstrom bond; glue thicknesses are estimated to be on the order of 1 $\mu m$. The Zr samples are then backed with optically transparent 6 mm thick [100] LiF crystals using similar bonding characteristics. The LiF serves as a tamper which maintains the pressure at this interface and allows for a measurement of the unloading response and phase reversion. On the opposite side of the gap, the LiF is bonded directly to the electrode. In both cases, a 3 mm diameter Al spot coating $\sim$1 nm thick is deposited on the bonded side of the LiF giving a reflecting surface for the VISAR \cite{Barker1972} diagnostic. VISAR provides the velocity-time history at each of these interfaces and is the primary experimental observable.

The configuration shown in Figure \ref{fig_RampConfig} is ideal for performing high-fidelity simulations, which is key for the interpretation in this work. The cathode measurement represents the velocity at the Al/LiF interface and is referred to as the \emph{drive} measurement because it allows for direct quantification of the magnetic pressure applied to the electrodes. This quantification is known as an \emph{unfold} \cite{Lemke2003} and consists of solving for the pressure-time history such that 1-D hydrocode simulations of the drive configuration reproduce the measured velocity. Thus, it is assumed the Al and LiF are well known standards; the material models are described in detail in ref. \onlinecite{Brown2014}. Conventionally, unfolds are performed through magneto-hydrodynamics (MHD) simulations to solve for the magnetic field, but for this experiment the magnetic effects are negligible and a pressure boundary condition is sufficient. The pressure drive is preferred here for simplicity and compatibility with the research code containing phase transition kinetics model. As suggested by the 1-D computational domain in Figure \ref{fig_RampConfig}, the benefit of this symmetric experimental configuration is the magnetic pressure must be the same across the electrode gap for a fixed height. Thus, once the pressure is determined for each drive measurement the only unknown in a simulation of the sample side is the Zr material model. These Zr models and their parameterizations are described in subsequent sections.

The shock experiment discussed below was carried out on a 50~mm gas gun.
A z-cut sapphire impactor struck a target consisting of a z-cut sapphire
buffer, the Zr sample, and a LiF window. The velocity signal at the Zr/LiF
interface was obtained from with a VISAR probe. A PDV probe measured the 
impactor  velocity. The shock breakout from the sapphire buffer was 
detected using a VISAR probe and used to infer the impact time.

The samples used here are very high purity Zr, with impurity levels 
in ppm by weight of: Hf 35, Fe $< 50$, Al $< 20$, V $< 50$, 
O $< 50$, N $< 20$, C 22. The material used here is the same as that 
designated Zr$_0$ in ref. \onlinecite{prigg_zr_jap}.

\section{Models}

The wave profile data are compared to forward simulations using a
multi-phase equation of state together with a phase transition kinetics model, 
which allows for the phase transitions to occur at finite rates.
A detailed description of the kinetics model is given in 
ref. \onlinecite{greeff_dyn_phase_2016}. Briefly, pressure and temperature
equilibrium among the coexisting phases is assumed, and the time evolution
of the phase fractions, ${\lambda_i}$ is given by,
\begin{equation}
\dot{\lambda_i} = \sum_{j \ne i} \left( \lambda_j R_{ji} - \lambda_i R_{ij} \right) \, .
\label{lambda_dot}
\end{equation}
This equation preserves the normalization of the phase fractions
$\sum_i \lambda_i =1$, and leads to asymptotic approach to complete
transformation. The functions $R_{ij}$ give the rate of transformation between
phase $i$ and $j$, and depend on the thermodynamic state. We use
the form\cite{greeff_ti_kinetics,greeff_ti_zr_sccm,greeff_dyn_phase_2016}
\begin{equation}
R_{ij} = \theta(G_i-G_j) \frac{\nu_{ij}}{B_{ij}} (G_i-G_j) 
   \exp[(G_i-G_j)^2/B_{ij}^2]  \, ,
\label{rfn_std}
\end{equation}
where $G_i$ denotes the Gibbs free energy of phase $i$ and $\nu_{ij}$
and $B_{ij}$ are the rate prefactor and energy scale respectively
for the $i\rightarrow j$ transition. They are used here as 
empirical parameters. Here $\theta(x)$ denotes the Heaviside step function.

A model for phase transition dynamics based on the physics of nucleation 
and growth has been successfully applied to solidification of water and
Ga.\cite{myint_h2o_frz_2018,myint_ga_solidification_2020} The case
of solid-solid transitions considered here is  not as well understood.
We use a phenomenological model
to infer information about transition rates from data, which may be helpful in
the development of more physics-based models.

We have found that some 
strain rate dependence of the flow stress is needed to match the 
observed rise of the
plastic wave in flyer impact experiments. Here we use the model due to
Swegle and Grady, \cite{swegle_grady_85} which has a well-defined yield
stress, above which the plastic strain rate varies as a power law in
the deviatoric stress. In the present uniaxial strain case, the model
takes the form
\begin{equation}
\dot{\epsilon}^p_{zz}  =  \theta\left(3 |s_{zz}|
           -2 Y_0 \right)
            C \frac{s_{zz}}{|s_{zz}|} \left( \frac{3 |s_{zz}|}{2 Y_0}
           -1 \right)^n
\end{equation}
where the wave propagation is the in the $z$ direction, $\dot{\epsilon}^p$
is the plastic strain rate, and $s$ is the deviatoric stress. The material
parameters are taken to have the values $C=1$ $\mu s^{-1}$, $n=2$,
$Y_0 = 0.4$ GPa, and the shear modulus is taken to be 36 GPa.

\section{Equation of State}\label{eos_section}
The equations of state are specified by giving the
Helmholtz free energies $F^{\sigma}(V,T)$ for each phase, $\sigma$. 
In this work, we take the parameterized free energies for $\alpha, \omega$,
and $\beta$ Zr described in ref. \onlinecite{greeff_zr_eos} as the starting 
point. There, the free energies were written as 
\begin{equation}
F^{\sigma}(V,T) = \phi^{\sigma}_0(V) + F^{\sigma}_{\rm ion}(V,T) 
           + F^{\sigma}_{\rm el}(V,T) \, ,
\label{f_total}
\end{equation}
where $\phi^{\sigma}_0$ is the static lattice energy, $F^{\sigma}_{\rm ion}$ 
and $F^{\sigma}_{\rm el}$ are the 
ion motion and electronic excitation free energies, respectively.
The static lattice energy was taken to have the Vinet form, \cite{vinet}
the ion motion term has the Debye form, and the the electronic excitation
free energy is 
$F^{\sigma}_{\rm el}(V,T) = - \frac{1}{2} \Gamma^{\sigma}(V) T^2$, 
corresponding to an electronic specific heat
$c_{V \rm{el}} = \Gamma^{\sigma}(V) T$.
Details of the volume
dependence of the Debye temperature and $\Gamma$ are given in 
ref. \onlinecite{greeff_zr_eos}.

The EOS has an equilibrium $\alpha-\omega$ transition at room temperature
and 2.2~GPa. This was built in as a constraint on the parameters, based
on the determination by 
Zilbershteyn {\em et al.}\cite{zilbershteyn_aw_ti_zr_73}.
They inferred the equilibrium transition pressure based on the fact that
under torsion, the forward and reverse transitions occurred at the 
same pressure. This interpretation was questioned by  Pandey and 
Levitas\cite{pandey_strain_induced_2020}, who prefer the value 3.4~GPa, 
obtained by extrapolating the high temperature phase boundary of 
Zhang {\em et al.}\cite{zhang_zr_phase_diag} to room temperature. 
Changing the equilibrium phase boundary in the EOS by this amount
would give different numerical values for the optimal kinetic parameters, 
but the qualitative picture would remain largely unchanged. The
$\alpha \rightarrow \omega$ transition occurs well above the equilibrium
pressure under dynamic compression, and this non-equilibrium is an 
important source of energy dissipation.

Here we make small modifications to the EOS from ref. \onlinecite{greeff_zr_eos}
for the $\alpha$ and $\beta$ phases. In both cases, these consist of changing 
parameters of the static lattice energy $\phi_0$. In 
reference \onlinecite{greeff_zr_eos}, the parameters of 
$\phi^{\alpha}_0(V)$ for the $\alpha$ phase were determined empirically
using the data of Fisher {\em et al.}, which gave the pressure derivative
of the bulk modulus as $dB_S/dP = 4.08$. More recent data from Liu {\em et al.}
gives\cite{liu_zr_ult_08} $dB_S/dP = 3.0$. We have modified the parameters
of $\phi^{\alpha}_0(V)$ to bring the $\alpha$ phase EOS into agreement with 
the Liu data. This also substantially improved agreement with density
functional theory (DFT) calculations using the PBE\cite{pbe} 
exchange-correlation functional,
which were described in ref. \onlinecite{greeff_zr_eos}. 
Recent static compression data
from Dewaele {\em et al.}\cite{dewaele_a-w_xafs_16} gives $dB_T/dP = 2.92$
based on fitting their room temperature $P-V$ data. 
Because this change is supported by independent data sets and theoretical
calculations, we view it as well-founded.

\begin{figure}
\noindent
\resizebox{1.0\columnwidth}{!}
{\includegraphics{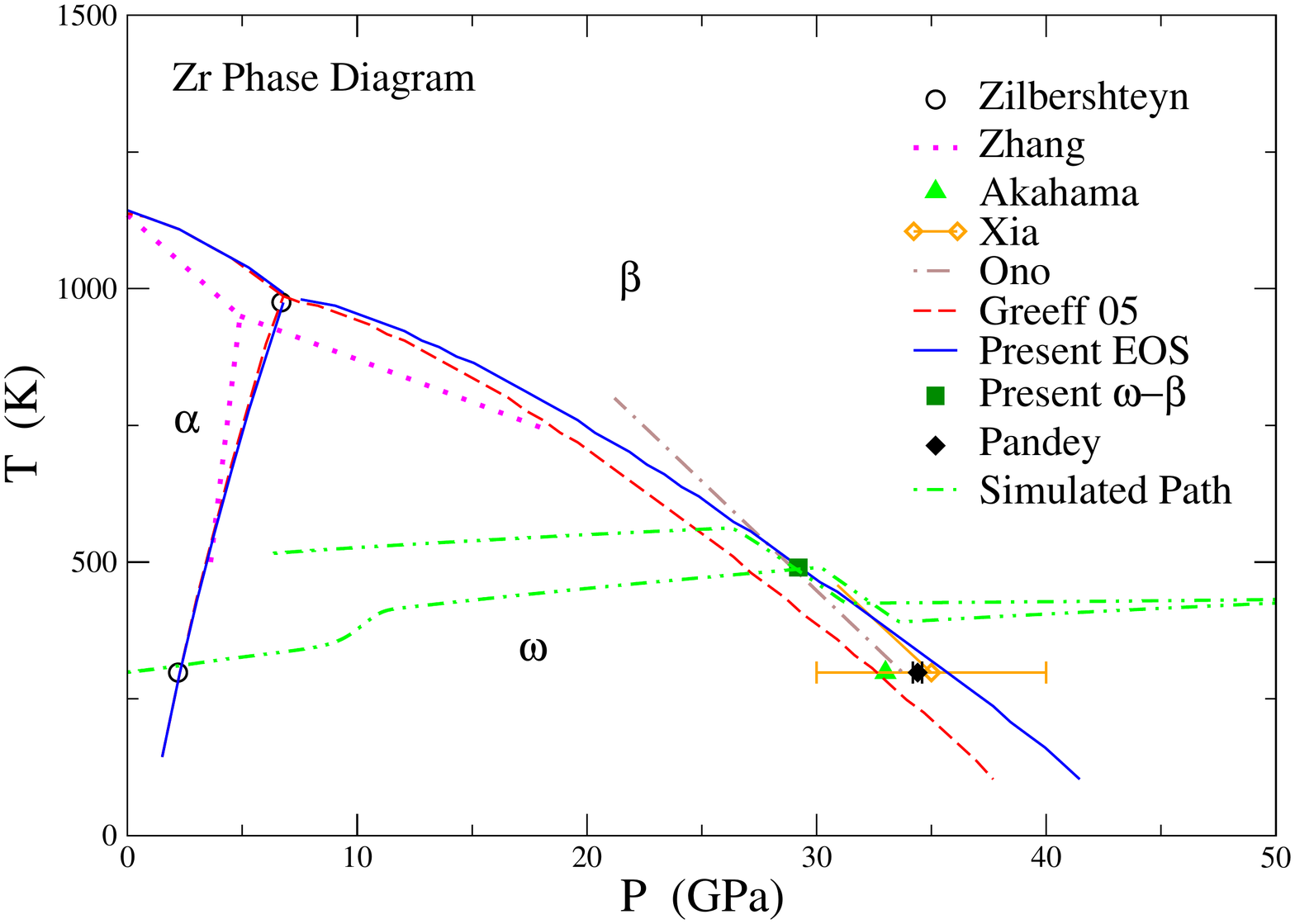}}
\caption{(color online) Phase diagram of Zr. Solid blue curves are phase 
boundaries of present EOS. Dashed red curves are from the EOS of 
ref. \onlinecite{greeff_zr_eos}. Data from 
refs. \onlinecite{zilbershteyn_aw_ti_zr_73,zhang_zr_phase_diag,
akahama91,xia90,xia91,ono_omega_beta_15,pandey_strain_induced_2020}. 
	Solid green square is present
estimate for intersection of ramp compression trajectory with equilibrium
$\omega-\beta$ boundary. Green dot-dot-dashed curve is simulated 
path of present ramp compression experiments.}
\label{phase_diag}
\end{figure}

In addition to the change made to the $\alpha$ phase EOS, we have modified the
$\beta$ phase EOS so as to increase the $\omega-\beta$ transition pressure.
It was found that this could be accomplished with minimal effect on
other properties by slightly increasing the cold bulk modulus of
the $\beta$ phase.  This change is made on the basis of the present ramp
compression data, which show both the $\omega \rightarrow \beta$ transition
on compression and the $\beta \rightarrow \omega$ transition on decompression.
It was not possible to get the transition in the right place in the forward and
reverse directions with the phase
boundary as it was placed in the original EOS, regardless of the 
kinetic parameters.
The value of the $\beta$ phase cold bulk modulus was chosen by 
simultaneously optimizing it 
and the kinetic parameters, as described below.
The resulting phase diagram is shown in figure \ref{phase_diag} along
with various data. \cite{zilbershteyn_aw_ti_zr_73,zhang_zr_phase_diag,
akahama91,xia90,xia91,ono_omega_beta_15} Our estimate for the intersection
of the ramp compression trajectory with the equilibrium $\omega-\beta$
line is shown as the solid green square. 

\section{Data and Analysis}
\begin{figure*}
\centering
	\includegraphics[width=\columnwidth]{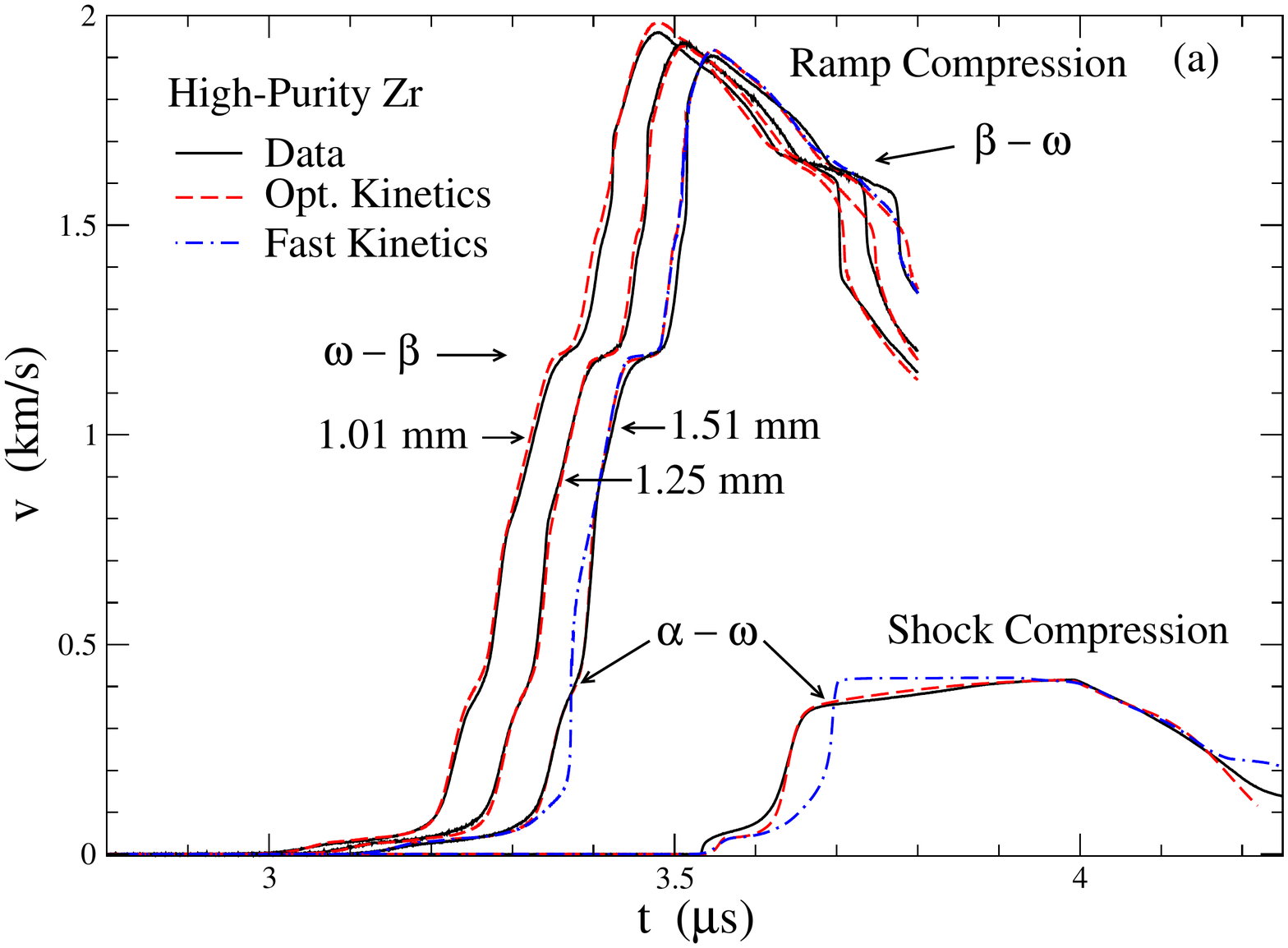}
\includegraphics[width=\columnwidth]{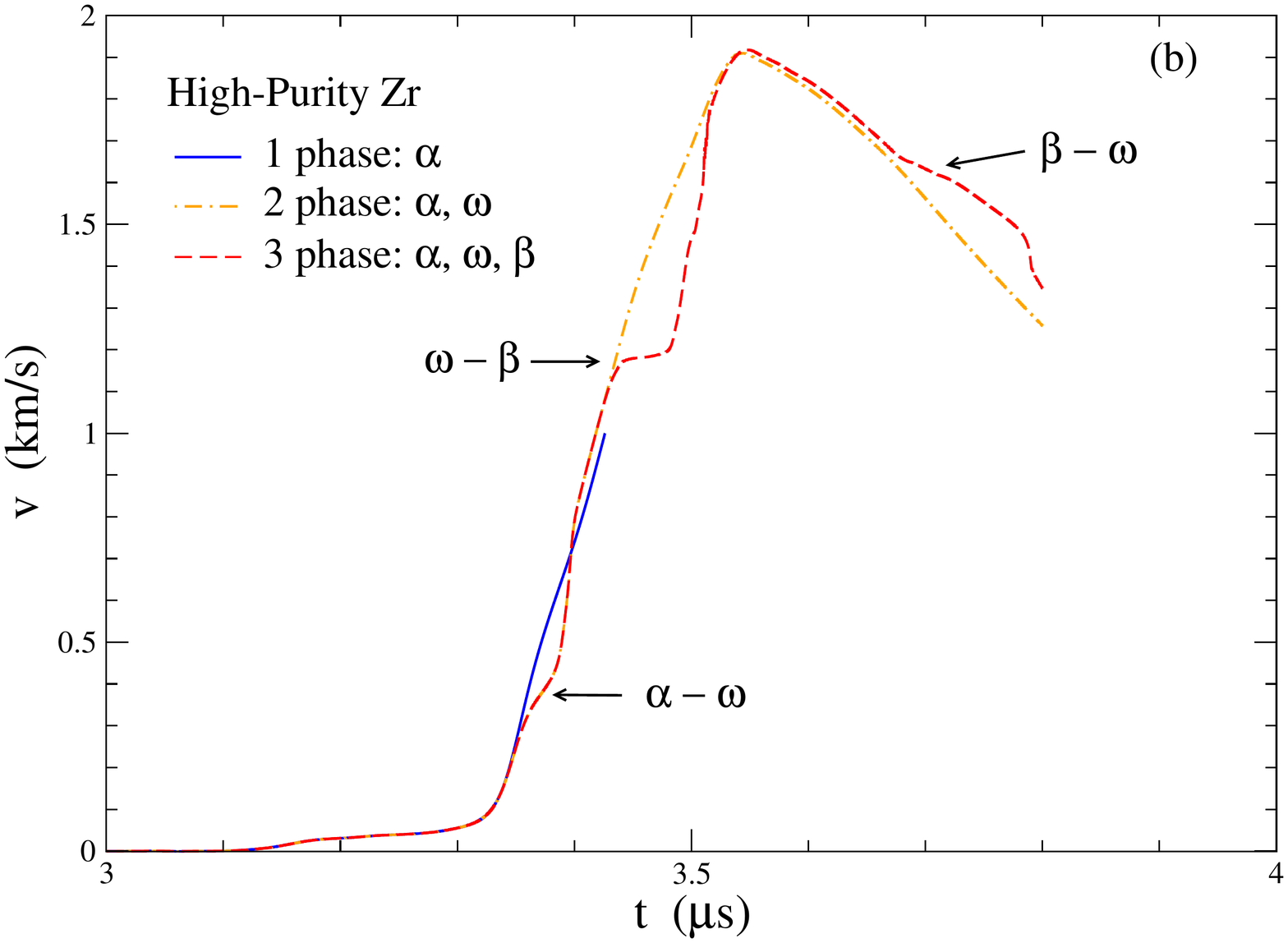}
	\caption{(color online) (a) Data and simulations for ramp and shock compression
of high-purity Zr, shot Z2913. Solid black curves are data, dashed red curves
are simulations with optimized kinetic parameters. Dot-dashed blue curves
	are simulations with fast kinetics that stay close to equilibrium. (b)
	Simulations with reduced number of phases to highlight the signatures of phase transitions. Solid blue curve is $\alpha$ phase only, dot-dashed orange curve is $\alpha$ and $\omega$ phases, red dashed curve is all three phases. Optimized kinetic parameters are used in every case.}
\label{highp_shock_ramp}
\end{figure*}

Figure \ref{highp_shock_ramp} (a) shows data for ramp and shock compression
experiments on high-purity Zr, along with simulations using our kinetic
model. In all cases the data consists of the velocity history at the
sample window interface. The ramp compression experiment, Z2913, 
consisted of Zr samples
with thicknesses of 1.01, 1.25, and 1.51~mm on Al electrodes with LiF
windows. The peak stress generated in the Zr was 56 GPa. The shock 
experiment, designated number 56-11-53\cite{rigg_l2_2015},
consisted of a sapphire impactor striking a sapphire buffer with the 
Zr sample and a LiF window attached. 
The Zr sample was 2.95~mm thick
and the impactor velocity was 0.54~km/s, producing a peak stress of 
9 GPa in the Zr. The earliest parts of the wave in both the shock and ramp
cases are elastic. The elastic wave is most easily distinguished in the shock
case. In figure \ref{highp_shock_ramp} (a), the earliest part of the shock 
velocity profile, with velocities $< 0.25$~km/s is an elastic wave, which is
followed by a slower plastic wave and subsequent phase transition. The ramp compression experiment shows the $\alpha \rightarrow \omega$, 
 the $\omega \rightarrow \beta$, and the reverse $\beta \rightarrow \omega$ transitions. 
The shock experiment shows the $\alpha \rightarrow \omega$ transition. This shock 
experiment is especially
sensitive to kinetics because at this pressure, the phase transition takes
~0.3 $\mu$s to complete, leading to the gradual rise of the velocity. The dashed red curves show
simulations using kinetic parameters optimized for the ramp compression
data. 

In order to highlight the features associated with phase transitions,
Figure \ref{highp_shock_ramp} (b) shows simulated velocity profiles
for the thickest (1.51 mm) ramp compression sample. The different simulations
incorporate varying numbers of phases from the $\alpha$ phase only
(solid blue curve), $\alpha$ and $\omega$ only (dot-dashed orange curve),
to all three $\alpha$,  $\omega$ and $\beta$ phases (dashed red curve). 
In all cases, the optimized kinetic parameters have been used.
On the rising side of the wave, each phase transition appears as a
plateau, associated with the low effective sound speed in the mixed-phase
region, followed by a steep rise reflecting the rapid increase in the sound
speed on completion of the transition. Similarly, on the decreasing side 
of the wave, the reverse $\beta \rightarrow \omega$ transition leads to a
plateau followed by a rapid drop in the velocity. 
Ref. \onlinecite{prigg_zr_jap} gives a more detailed discussion of wave features
in relation to EOS and phase transitions.

Also shown in Figure \ref{highp_shock_ramp} (a) as the dashed 
curves are simulations using fast kinetics,
which are essentially in equilibrium. Under equilibrium conditions, the 
ramp compression experiment forms a shock, as indicated in the figure by
the rapid velocity increase between 0.18 and 0.58 km/s. In the fast kinetic
simulations, the sample transforms directly 
to the $\omega$
phase in this shock. The shock experiment similarly transforms 
completely to the $\omega$
phase in the plastic wave. In equilibrium there is no gradual rise in the
velocity, as seen in the data, and the plastic wave is too slow.

\begin{figure}
\noindent
\resizebox{1.0\columnwidth}{!}
{\includegraphics{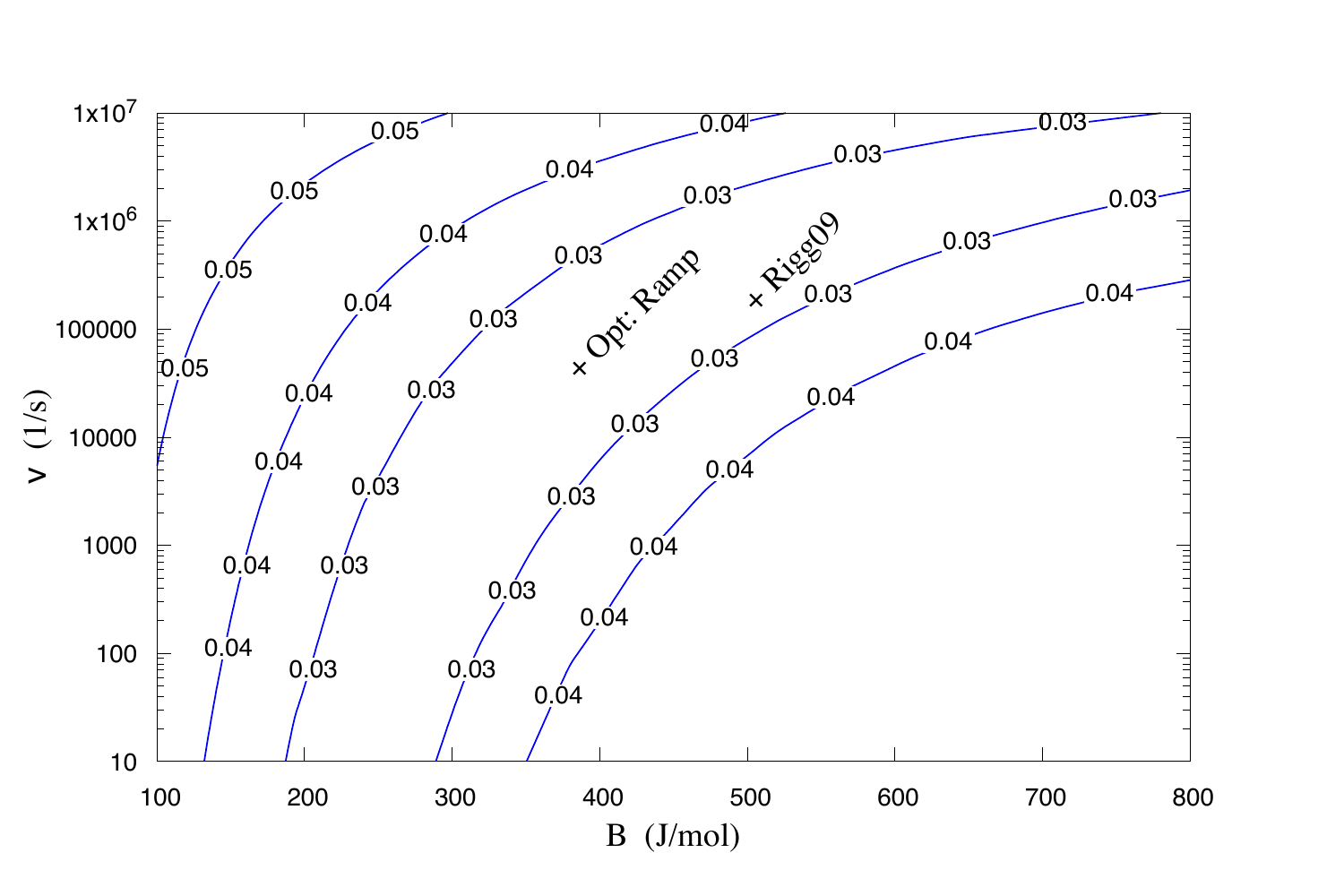}}
\caption{(color online) Contours of the RMS error in the wave profile
as a function of the kinetic parameters $\nu$ and $B$ for the $\alpha-\omega$
transition in high purity Zr. The contours are for the ramp 
compression experiment Z2913.
The point labeled ``Opt: Ramp'' is the minimum error for this experiment
and the one labeled ``Rigg09'' was approximately optimized by hand for
shock experiments\cite{prigg_zr_jap}.} 
\label{highp_contour}
\end{figure}

In past applications of the kinetics model\cite{greeff_ti_zr_sccm,prigg_zr_jap}
we have determined approximately optimal kinetic parameters by hand.
In this work we have optimized parameters by minimizing the rms error
of the velocity profile
\begin{equation}
E^2 = \frac{1}{N} \sum_{i=1}^N \frac{1}{t_2-t_1} \int_{t_1}^{t_2} dt \left[ v_i^{\rm sim}(t) - v_i^{\rm exp}(t) \right]^2
\end{equation}
The index $i$ denotes different data sets. Here it refers to different sample
thicknesses. The $\alpha-\omega$ and $\omega-\beta$ parameters affect different
parts of the
wave profile, so they have been independently optimized here.
The present optimization algorithm scans over a grid that is uniform in $B$
and logarithmically uniform in $\nu$ and finds the minimum rms error
on the grid. 

In the case of the $\omega-\beta$ transition, the 
equilibrium phase boundary needs to be adjusted to allow for a good 
match to the wave profile on both the forward and reverse transitions. 
As described in section \ref{eos_section}, the phase boundary was adjusted
using the cold bulk modulus of the $\beta$ phase as a parameter. The kinetic
parameters for the forward $\omega \rightarrow \beta$ and reverse 
$\beta \rightarrow \omega$ transitions affect different parts of the wave
profile, and were independently optimized by scanning $(B,\nu)$
grids as for the $\alpha \rightarrow \omega$ transition.
The kinetic parameters were optimized for a sequence of values of
the $\beta$ phase static lattice bulk modulus, starting from the value 
79.2 GPa of the 2015 EOS\cite{greeff_zr_eos}, increasing in steps
of 0.65 GPa. The minimum error occurred at a value of 80.5 GPa.  
For the $\omega-\beta$ transition, optimization favors small
values of the kinetic parameter $B$. However if $B$ is too small, 
the simulations experience numerical
instabilities. We have set a lower cutoff of 50 J/mol for $B$. The 
minimum error for the forward $\omega \rightarrow \beta$ transition 
was found at $B = 50$ J/mol and $\nu = 10$ s$^{-1}$, while for the
reverse $\beta \rightarrow \omega$ transition, the best values are 
$B = 50$ J/mol and $\nu = 10^{7}$ s$^{-1}$. These parameter values
bring both transitions close to equilibrium, so there is very little
difference in the wave profiles from the equilibrium case for the
$\omega-\beta$ transition in figure \ref{highp_shock_ramp}.

Figure \ref{highp_contour} shows error contours in the $B_{\alpha \omega},
\nu_{\alpha \omega}$ plane for the ramp compression experiment Z2913
on high-purity Zr. The symbols mark the optimum value for Z2913,
$B_{\alpha \omega} = 380$~J/mol, 
$\nu_{\alpha \omega} = 3.98 \times 10^4 {\rm s}^{-1}$  and the
parameters used in Rigg {\em et al.} \cite{prigg_zr_jap} for shock loading
experiments 
$B_{\alpha \omega} = 500$~J/mol,
$\nu_{\alpha \omega} = 1.7 \times 10^5{\rm s}^{-1}$.
 While the parameter values are rather different, the rms errors associated
with the parameter pairs are very similar. There is strong correlation between
$B$ and $\nu$, and the error  surface has a long valley, which is very
flat. The rms error changes by only ~1\% while $B$ varies from 240 to 
473 J/mol and $\nu$ varies from 10 to $3.6 \times10^5$ s$^{-1}$.  Because 
$B$ appears in the exponential, while $\nu$ is outside it, $\nu$ varies
by several orders of magnitude over this range of $B$. 

\begin{figure}
\noindent
\resizebox{1.0\columnwidth}{!}
{\includegraphics{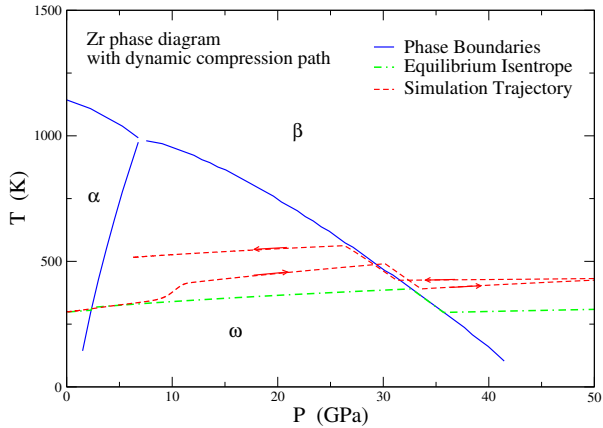}}
\caption{(color online) Simulated thermodynamic path of 
ramp compression experiment Z2913. Solid blue curves are the equilibrium phase 
boundaries of the current EOS. Green dot-dashed curve is the equilibrium 
isentrope. Red dashed curve is the simulated path of an interior point of 
the sample, with arrows indicating increasing time. This differs from the 
isentrope due to both plastic dissipation and non-equilibrium phase 
transitions.}
\label{phase_diag_dynamic_comp}
\end{figure}

Figure \ref{phase_diag_dynamic_comp} shows the $P,T$ trajectory from our 
simulation of ramp compression of experiment  Z2913 with the equilibrium 
phase boundaries from our EOS. Also shown is the isentrope $S= {\rm const}$.
In the idealization of no dissipation, the trajectory would follow
the isentrope. There are two sources of dissipation in the
simulation: plastic work, and non-equilibrium phase transitions. 
On the compression path, the material remains largely in the 
$\alpha$ phase until the pressure reaches 7 GPa, and is nearly completely
transformed to the $\omega$ phase at 12 GPa. 
This interval corresponds
to a temperature rise where the simulation trajectory departs from
the isentrope. The $-P dV$ work on this non-equilibrium path is larger than 
on the isentrope, resulting in dissipative heating. At 29.4 GPa, where
the 
simulation path crosses the equilibrium $\omega-\beta$ boundary, it lies 
100~K above
the isentrope. By comparing simulations with no strength or equilibrium kinetics with nominal models, we
estimate that about half of this temperature rise is due to plastic work,
and half is due to the non-equilibrium $\alpha-\omega$ transition.
The transformation completes at a higher pressure in the ramp case 
than the shock case because, under ramp
compression, the pressure rises continuously as the transformation proceeds, 
whereas in the shock case, the pressure is nearly constant for $\sim 0.4 \mu$s,
allowing time for completion at a lower pressure. At the same time, the 
transformation rate increases rapidly with pressure, so the overall time 
for the transformation is shorter in the ramp case.

\begin{figure}
\noindent
\resizebox{1.0\columnwidth}{!}
{\includegraphics{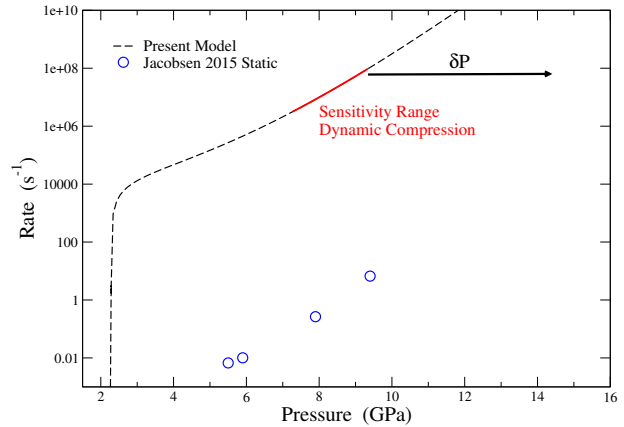}}
\caption{(color online) Phase transition rate in the present model,
Eq. (\ref{rfn_std}) with optimized parameters (black dashed curve) compared with 
observed rate in DDAC experiments\cite{jacobsen15}. The red segment of the curve
shows the range of rates that affect the simulated velocity profile
in ramp compression experiment Z2913. The arrow labeled $\delta P$
shows the estimated pressure equivalent to the shear stress as it influences
the nucleation rate.}
\label{rate_static_dynamic}
\end{figure}

The time evolution of the $\alpha-\omega$ transition was observed via
diffraction by Jacobsen {\em et al.} \cite{jacobsen15}. These measurements use
a dynamic diamond anvil cell (DDAC) apparatus. After pre-compressing within the
$\alpha$ phase, a piezoelectric module applied a step increase in the
pressure over a time of $< 0.1$~s. The $\omega$ phase
fraction $\lambda_{\omega}$ was obtained following the pressure jump at 
room temperature. The data were analyzed to extract a time constant
$\tau$, which is the time required for $\lambda_{\omega}$ to reach
$1 - e^{-1}$. If the present kinetic model is applied to the same situation, the
corresponding time is $R_{\alpha \omega}^{-1}$. It is therefore 
meaningful to compare $R_{\alpha \omega}$ to $1/\tau$, which is done
in figure \ref{rate_static_dynamic}. The open circles are DDAC data from
Jacobsen {\em et al.} and the dashed black curve is $R_{\alpha \omega}$,
evaluated from Eq. (\ref{rfn_std}) along the room temperature isotherm
with optimized parameters. 
The solid red segment of the model curve indicates the range over which
the present dynamic compression simulations are sensitive 
to $R_{\alpha \omega}$. The lower limit was determined by carrying out a 
series of simulations with the rate $R$ set to zero if it fell below
a threshold. For values below approximately 
$3 \times 10^{6}$~s$^{-1}$,  this threshold made no noticeable difference 
to the simulated wave profile, whereas above this there was a significant
change. The upper end corresponds to the highest rates in our 
simulations, which  were $\sim 10^8$~s$^{-1}$. 

Because of this limited sensitivity range, the functional form 
Eq.(\ref{rfn_std}) is not unique, and any function giving a linear dependence
of $\ln R_{ij}$ on the driving force $G_i - G_j$ would give similar results.
For example, it was found in ref. \onlinecite{greeff_dyn_phase_2016} that
the form 
$R_{ij} = \theta(G_i-G_j) \xi_{ij} \sinh\left[(G_i-G_j)/C_{ij}\right]$
gives nearly indistinguishable velocity profiles to Eq.(\ref{rfn_std}),
when the parameters $\xi$ and $C$ are determined so as to 
match Eq.(\ref{rfn_std}) in the sensitivity range.

Figure \ref{rate_static_dynamic} shows that the 
sensitivity range for dynamic compression overlaps in pressure with the
DDAC measurements. In this overlapping range, the
transition rate under dynamic compression exceeds that under quasi-static 
compression by a factor of $\sim 3 \times 10^7$. The temperature
is somewhat higher in the dynamic compression case, ranging from 350-400~K
at the time of peak transformation rate. Experimental estimates of
the activation energy of the transformation are in the range 0.5-1.73 eV
\cite{jacobsen15,zong2014}. Taking the smallest activation energy and 
and largest $T$ for the dynamic compression experiments leads to a factor
of $1.3 \times 10^2$ between the rates, so it is unlikely that the 
temperature accounts for the observed difference. A temperature of 
$2.3 \times 10^3$~K would be required to account for the rate difference.

It is well established that shear stress and shear deformation 
strongly influence the $\alpha-\omega$ transition in Ti and Zr.
\cite{zilbershteyn_aw_ti_zr_73,errandonea2005}
A possible mechanism for this influence is through a change in the
nucleation rate by shear stress. The rate of nucleation is proportional 
to $e^{-W^*/k_b T}$, where $W^*$ is the free 
energy of a critical nucleus, which is in turn a function of the bulk free
energy difference between the daughter and
parent phases. 
This exponential dependence of the nucleation rate on the free energy 
difference provides a natural explanation for the exponential 
dependence of our phenomenological rate, Eq. (\ref{rfn_std}), if
nucleation is the limiting process. 

A model for the influence of shear stress on the nucleation rate of 
a martensitic transition was proposed by Fisher and Turnbull.\cite{fisher1953}
They considered the case of a thin, lenticular second phase domain,
with the transformation strain taken to be a simple shear,
$\epsilon_{xy} = \theta$, under the assumption of a coherent interface. 
They modeled the influence of a shear
stress $s_{xy} = \tau$ and found that that its effect on $W^*$ is 
to replace the bulk free energy difference $\Delta g = \Delta G/V$ with 
\begin{equation}
\Delta g \rightarrow \Delta g - \frac{4}{3} \tau \theta \, .
\label{stress_g}
\end{equation}
Noting that $\tau \theta$ is the work $w$ per unit volume done by
the applied shear stress on the transforming domain, we generalize this as 
\begin{equation}
\Delta g \rightarrow \Delta g - c w = 
            \Delta g - c \epsilon_{ij} s_{ji}
\label{gen_stress_g}
\end{equation}
where $\epsilon_{ij}$ is the transformation strain, $s_{ji}$ is
the deviatoric stress, and $c$ is a geometric factor of order unity
that is related to the shape of the second phase domain. The factor $c$
differs from unity because of the strain energy in the parent phase
matrix, and because the strains within the daughter phase domain will
differ from the ideal transformation strains $\epsilon_{ij}$. Linearizing
$\Delta g$ with respect to $P$, we find that the shear stress has the same
effect on the nucleation rate as an additional pressure
\begin{equation}
\delta P = \frac{c \epsilon_{ij} s_{ji}}{\Delta V/V} \, .
\label{stress_P}
\end{equation}

Consider, for example, the TAO-1 mechanism for the $\alpha-\omega$ 
transition\cite{trinkle_2003} with transformation strains $\epsilon_{xx}=-0.09$,
$\epsilon_{yy} = 0.12$, and $\epsilon_{zz} = -0.02$, in the standard hcp
crystal axes.
For the case of uniaxial compression, the macroscopic shear stress is 
of the form,
\begin{equation}
\left( \begin{array}{ccc}
-s^\parallel/2 & 0 & 0 \\
0 & -s^\parallel/2  & 0 \\
0 & 0 & s^\parallel
\end{array} \right)
\end{equation}
in a frame with the $z$-axis aligned with the propagation direction.
The shear stress enhancement is maximized
when the compression wave propagates in the crystal $x$-direction, giving
$w = \epsilon_{ij} s_{ji} = 0.14 |s ^{\parallel}|$, where $s ^{\parallel}$ 
is the deviatoric stress in the wave propagation direction.
The fractional volume
change is $\Delta V/V = 0.01$ for the Zr $\alpha-\omega$ transition,
and our simulations give $|s ^{\parallel}| = 0.5$~GPa during the transition.
So the shear stress enhances the nucleation rate by the same amount as
an additional pressure $\delta P \sim 7$~GPa. This estimate corresponds
to the TAO-1 mechanism with the optimal orientation of the crystal
with respect to the propagation direction. Polycrystalline samples will
sample a distribution of orientations, and other mechanisms with different
transformation strains may be active. Accounting for this, we expect
a range of $\delta P$ on the order of several GPa. The small value
of $\Delta V/V$  in the denominator of Eq. (\ref{stress_P}) amplifies the
effect of shear stress.

Because our rate model does not explicitly account for shear stress, but 
is calibrated to data in which it is present, the hydrostatic rate function
will be shifted to the right by $\delta P$ with respect to the curve in
figure \ref{rate_static_dynamic}. A shift of several GPa, as suggested by 
the above analysis, will bring  the model into better alignment with 
with the extrapolated DDAC 
compression data of Jacobsen {\em et al.} \cite{jacobsen15}. This is 
illustrated in figure \ref{rate_static_dynamic} with the arrow, whose length
is 5~GPa. Most of the DDAC experiments were done without a pressure 
transmitting medium, and were not fully hydrostatic. The shear stress
was not quantified in those experiments, but, given their much lower rate,
it is expected to be lower than that of the current dynamic experiments.

\section{Conclusions}

We have presented new data on ramp compression of high purity Zr that
show the $\alpha-\omega$  and $\omega-\beta$ phase transitions, with the
higher pressure $\omega-\beta$ transition occurring in the forward direction on
compression and  reversion direction on release. Simulations employing
a multi-phase equation of state and a phenomenological kinetic model
match the experimental velocity profiles well. The same parameters also
agree well with shock compression data on the $\alpha-\omega$ transition.
The data showing both the forward and reverse $\omega-\beta$ transitions 
allows us to simultaneously optimize the kinetic parameters and parameters of
the EOS, enabling us to refine our estimate of the equilibrium
$\omega-\beta$ phase boundary. The resulting phase boundary is higher in
pressure than that of an earlier EOS \cite{greeff_zr_eos}.
We find that, under dynamic compression, the $\alpha-\omega$ transition 
overshoots the equilibrium phase boundary by $\sim 9$~GPa, while
the $\omega-\beta$ transition occurs much closer to equilibrium in both
the forward and reverse directions.

The $\alpha-\omega$ transition shows strong kinetic effects.
We  find that the wave profiles for these experiments are 
sensitive to phase transition rates in the range 
$3 \times 10^6 - 10^8$~s$^{-1}$. The requirement for the model to match data 
is that the 
logarithm of the rate depends approximately linearly on the 
thermodynamic driving force in this range. 
The non-equilibrium $\alpha-\omega$ transition is estimated to account
for half of the dissipative temperature rise of 100~K at the onset
of the high pressure $\omega - \beta$ transition.

Eq. (\ref{stress_P}), $\delta P = (c \epsilon_{ij} s_{ji})/(\Delta V/V)$, 
relates the shear stress, $s_{ji}$, to an equivalent pressure increase,
$\delta P$, as it 
influences the phase transition rate. 
In the present case, our analysis was motivated by 
a model for the nucleation rate.\cite{fisher1953} 
However, the derivation of Eq. (\ref{stress_P}) only involves the 
bulk free energies, so it is likely to be more generally valid.
In the case of the TAO-1 mechanism with optimal orientation 
considered above, $\delta P \approx 7$~GPa, while the shear stress is 0.5~GPa.
This amplification results from the factor $\epsilon/(\Delta V/V)$, where
in this case, the fractional volume change is small compared to the 
transformation shear strains. The present kinetic model gives transformation
rates several orders of magnitude larger than those observed in DDAC 
experiments\cite{jacobsen15} in an overlapping pressure range.  
Our estimate for the shear stress effect is approximately the right size to
explain the difference. However, the DDAC experiments were not fully
hydrostatic, so other mechanisms may be required to explain the difference.

\begin{acknowledgments}
	This work was supported by the U.S. Department of Energy through the Los Alamos National Laboratory. Los Alamos National Laboratory is operated by Triad National Security, LLC, for the National Nuclear Security Administration of U.S. Department of Energy (Contract No. 89233218CNA000001).
	NV work performed under the auspices of the U.S. Department of Energy by Lawrence Livermore National Laboratory under Contract DE-AC52-07NA27344.
\end{acknowledgments}

\bibliography{zdp_all}

\end{document}